\documentclass[letterpaper, 10 pt, conference]{ieeeconf}  

\IEEEoverridecommandlockouts                              
\usepackage[utf8]{inputenc}
\usepackage[T1]{fontenc}
\usepackage{amsmath,amssymb,amsfonts}
\usepackage{mathtools}
\usepackage{subfig}
\usepackage{graphicx}
\usepackage{siunitx}
\usepackage{array}
\graphicspath{{figures/}} 
\usepackage{amsmath}
\usepackage{mathtools,amssymb,lipsum}

\usepackage{cuted}
\usepackage{relsize}
\usepackage{caption}
\usepackage{multirow}
\usepackage{tabularx}
\usepackage{placeins}
\usepackage{float}
\usepackage{multicol}
\usepackage{dirtytalk}
\usepackage{upgreek}

\title{\LARGE \bf
Predictions of Electromotive Force of Magnetic Shape Memory Alloy (MSMA) Using Constitutive Model \& Generalized Regression Neural Network
}

\author{Md Esharuzzaman Emu$^{1}$ 
\thanks{$^{1}$Md Emu is a Graduate Student, Dept. of Mechanical Engineering,
        University of Illinois at Chicago, 842 West Taylor Street 
Chicago, IL 60607-7022
        }%
}

\begin{document}

\maketitle
\thispagestyle{empty}
\pagestyle{empty}

\begin{abstract}

Ferromagnetic shape memory alloys (MSMAs), such as Ni-Mn-Ga single crystals, can exhibit the shape memory effect due to an applied magnetic field at room temperature. Under a variable magnetic field and a constant bias stress loading, MSMAs have been used for actuation applications. Under variable stress and a constant bias field, MSMAs can be used in power harvesting or sensing devices, e.g., in structural health monitoring applications. This behavior is primarily a result of the approximately tetragonal unit cell whose magnetic easy axis is approximately aligned with the short axis of the unit cell within the Ni-Mn-Ga single crystals.  Under an applied field, the magnetic easy axis tends to align with the external field.  Similarly, under an applied compressive force, the short side of the unit cell tends to align with the direction of the force. 

This work introduced a new feature to the existing macro-scale magneto-mechanical model for Ni-Mn-Ga single crystal. This model includes the fact that the magnetic easy axis in the two variants is not exactly perpendicular as observed by D’silva et al. \cite{Silva}. This offset helps explain some of the power harvesting capabilities of MSMAs.

Model predictions are compared to experimental data collected on a Ni-Mn-Ga single crystal. The experiments include both stress-controlled loading with constant bias magnetic field load (which mimics power harvesting or sensing) and field-controlled loading with constant bias compressive stress (which mimics actuation). Each type of test was performed at several different load levels, and the applied field was measured without the MSMA specimen present so that demagnetization does not affect the experimentally measured field as suggested by Eberle et al. \cite{Eberle_2019}. Results show decent agreement between model predictions and experimental data.

Although the model predicts experimental results decently, it does not capture all the features of the experimental data. In order to capture all the experimental features, finally, a generalized regression neural network (GRNN) was used to train the experimental data (stress, strain, magnetic field, \& emf) so that it can make a reasonably better prediction.

\end{abstract}

\section{INTRODUCTION}

\begin{figure}[H]
      \centering
      \includegraphics[scale=0.5]{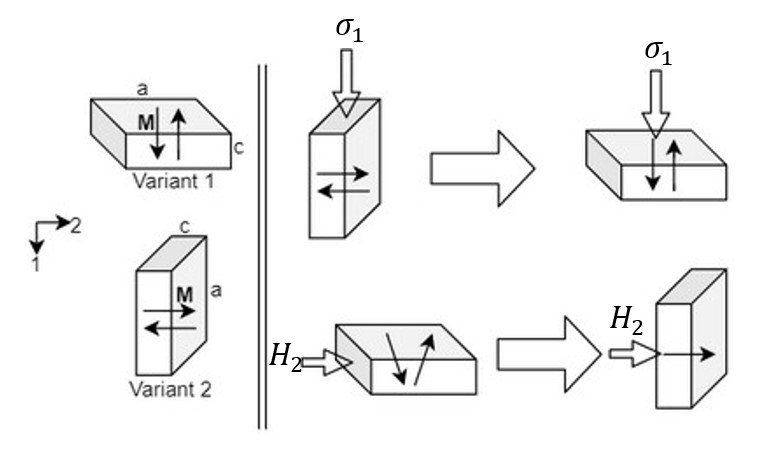}
      \caption{Left: Tetragonal variants in Ni-Mn-Ga samples. The solid arrows represent the internal magnetization vectors, which are approximately aligned with the short side of the unit cell. The direction of the short side (c) of the unit cell defines each variant. Right: Typical behavior of MSMA under applied stress (top) or field (bottom). Notice the short side of the unit cell tends to align with the direction of an external compressive stress or the direction of an external applied magnetic field.}
      \label{fig:1w}
   \end{figure}

Magnetic shape memory alloys (MSMAs) are a type of smart material that exhibits magnetic field and stress-induced shape memory effect at room temperature.  The most common alloy in this class is NiMnGa, which is the material used in this work. The twin variants in NiMnGa alloys reorient at low-stress levels, which changes the material’s magnetization. This phenomenon makes MSMAs suitable for power harvesting and sensing applications ~\cite{Guiel2018}. Also, the twin variants reorient under the applied field. This, along with the relatively fast response ($\approx$ 1 kHz) of MSMAs, makes these materials suitable for actuation~\cite{Minorowicz_2016} and micropump applications \cite{Ullakko2012}.
\\
The microstructure of the material shown in Fig.~\ref{fig:1w} can explain the unique behaviour. The magnetic easy axis is approximately aligned with the short side of the unit cell (Fig.~\ref{fig:1w} left). When a compressive stress is applied to the material, the short side of the unit cell tends to align with the direction of the stress (Fig.~\ref{fig:1w}, top, right). Once fully aligned, this configuration is called the stress preferred variant. The magnetization vector \textbf{\textit{M}} begins to align with the direction of the external field when magnetic field \textbf{\textit{H}} is applied (Fig.~\ref{fig:1w} right, bottom). Once fully aligned, this configuration is known as the field preferred variant. As can be seen in Fig.~\ref{fig:1w}, transitions between stress preferred and field preferred variants or vice versa causes a change in magnetization vector inside the material and the bulk magnetic field around it which can be harvested into useful voltage in conjunction with the proper setup \cite{Nelson2014}.
For proper implementation of MSMA as a sensor, energy harvester, and actuator, a generalized constitutive model is required that can predict the behavior of MSMA subject to general magneto-mechanical loading. Several constitutive models for the macro scale behavior of MSMAs have been proposed, including but not limited to, those by Kiefer and Lagoudas \cite{Kiefer2005}, Wang and Steinmann \cite{Wang2012}, LaMaster et al. \cite{LaMaster2014}, and others. But no model has been shown to satisfy all the laws of thermodynamics and accurately capture the magnetization and applied field (\textbf{\textit{M}} vs. \textbf{\textit{H}}) curves as well as the stress-strain and field-strain response under a wide array of loading conditions. Nelson et al. \cite{Nelson2014} attempted to predict the electromotive force (emf) generated by MSMA under certain loading by using LaMaster’s model \cite{LaMaster2014}. However, there was a noticeable difference between predicted peak to peak emf and experimental peak to peak emf. Further, D Silva et al. \cite{Silva} observed that the field preferred and stress preferred variants are not exactly perpendicular to each other. This offset from perpendicular has not been included in any models.

In order to predict mechanical behavior, besides using constitutive modeling, machine learning-based modeling is getting popular \cite{DesignedResearch;M2019,Cyron2021,Linka2021,Zhang2022}. Machine learning-based modeling, specifically neural networks can predict nonlinear behavior very well. In the present work, a general constitutive model is proposed that can predict the \textbf{\textit{M}} vs. \textbf{\textit{H}} curves correctly and give a decent prediction of the stress-strain and field-strain curve under various load conditions by using the manufacturer given anisotropy constant. The offset angle from perpendicular to the variants is also included in the model. After that, the model is used to predict emf. Finally, a generalized regression neural network (GRNN) \cite{97934} based model is used to predict the emf. The predictions of emf from both models are compared.

\section{Constitutive Modeling}
\begin{figure}[htbp]
\centering
\includegraphics[scale=0.5]{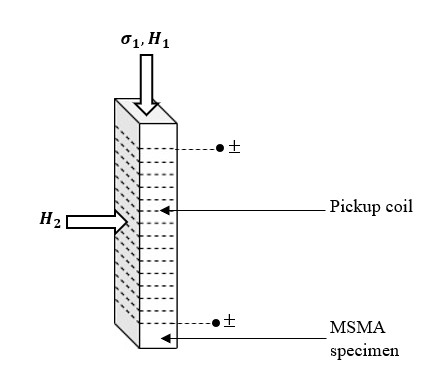}
\caption{Schematic of power harvesting setup with MSMA specimen warped by pickup coil under varying compressive stress $\sigma_1$, constant axial field $H_1$, and lateral field $H_2$.}
\label{fig:2w}
\end{figure}

Among the many potential applications of MSMA, power harvesting is one of the most promising ones. MSMAs are becoming popular in energy harvesting applications. Their high fatigue life around $10^9$ cycles~\cite{Tellinen2002}, ease of design, particularly compared to piezo and magneto strictive material~\cite{niskanen2013} and the range of operational frequencies, 2-100 Hz~\cite{niskanen2013}. In power harvesting applications, an MSMA specimen is exposed to variable mechanical stress in combination with a constant external magnetic field. During mechanical loading, the internal magnetization vector of the MSMA changes. This facilitates power generation in a pickup coil surrounded on MSMA specimen (see Fig.~\ref{fig:2w}). In the configuration shown in Fig.~\ref{fig:2w}, the MSMA specimen is kept under a constant lateral magnetic field $H_2$ (0.3-0.8 Tesla) with a varying compressive load of $\sigma_1$ between 0 to 5 MPa to generate emf. Previously Guiel et al.~\cite{Guiel2018} found that a small constant axial magnetic field $H_1$ (0.03-0.1 T) dramatically increased the emf output of the power harvesting system shown in Fig.~\ref{fig:2w}. In order to design and optimize such power harvesting devices, a model is required to predict emf under the loading conditions shown in Fig.~\ref{fig:2w}.

\begin{figure}[htbp]
\centering
\includegraphics[scale=0.5]{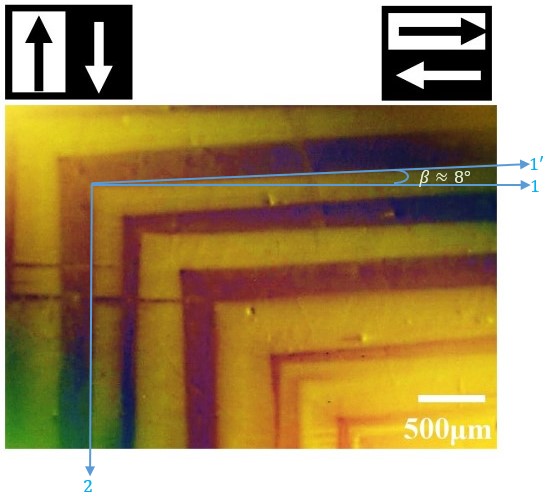}
\caption{Magnetic domain pattern in-plane showing magnetic domains in the field and stress preferred variant, as viewed through a Magneto-Optical Indicator Film (MOIF). Reproduced from \cite{Silva}.}
\label{fig:6w}
\end{figure}

\begin{figure}[htbp]
\centering
\includegraphics[scale=0.5]{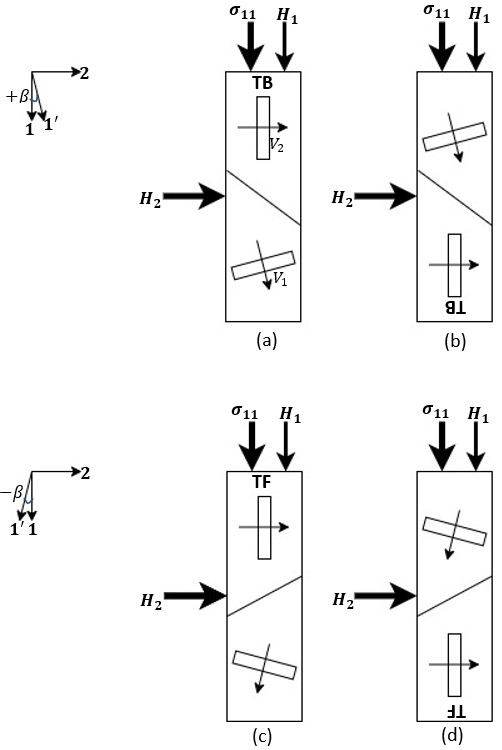}
\caption{All the possible orientations of the MSMA specimen. TF stands
for the top and front side of the specimen, TB stands for the top
and back side of the specimen. $V_1$ and $V_2$ represent stress preferred, and field preferred variants, respectively. The arrows indicate possible
orientations of the internal magnetization vector. The possible orientations include the offset angle $\beta$ in $V_1$. Reproduced from \cite{Nelson2014}.}
\label{fig:27w}
\end{figure}

In order to give physical grounding to this work, the predictions of emf will be based on the model by LaMaster et al.~\cite{LaMaster2014}. This model is based on sufficient conditions to satisfy the laws of thermodynamics and has been shown to be able to predict stress-strain experiments and some field-strain experiments reasonably well. However an offset angle observed experimentally by D Silva et al.~\cite{Silva} is introduced to the model. Light microscopy on MSMA surface reveals that the field preferred and stress preferred variants are not exactly perpendicular to each other (see Fig.~\ref{fig:6w}). The offset from perpendicular between the two variants will be called $\beta$. The value of $\beta$ can be arbitrarily positive or negative depending on how the specimen is oriented as shown in Fig.~\ref{fig:27w}. By performing the stress-controlled test, Nelson et al.~\cite{Nelson2014} experimentally showed that the orientations shown in Fig.~\ref{fig:27w} (a) $\&$ Fig.~\ref{fig:27w}(b) produced similar emf as well as Fig.~\ref{fig:27w}(c) $\&$ Fig.~\ref{fig:27w}(d) produced similar emf. Fig.~\ref{fig:27w}(a, b) and Fig.~\ref{fig:27w}(c, d) presume that $\beta$ is positive and negative, respectively. This offset angle may explain the different responses of the material with different orientations of the specimen as represented by Nelson et al.~\cite{Nelson2014}.

\subsection{Model Derivation}
\begin{figure}[H]
\centering
\includegraphics[scale=0.5]{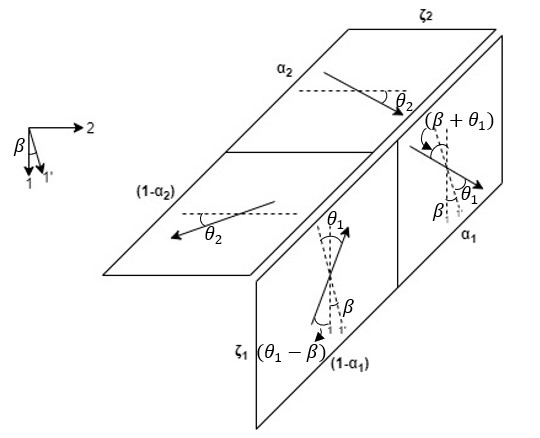}
\caption{Idealized microstructure with offset angle ($\beta$) added, assuming only two variants, distinct domain walls in each variant and magnetization vector rotation.}
\label{fig:7w}
\end{figure}

The model is homogenized~\cite{T_lance_2018}, which means that the model does not capture specific features of the microstructure but rather captures the average behaviour of the material at a point. In this case, it is also assumed that the load on the MSMA is uniform, and thus, all points in the material in the same sense. The model captures the average behaviour through three internal variables which are each associated with its microstructure: magnetization vector rotation ($\theta$), volume fraction of the magnetic domain ($\alpha$), and volume fraction of variants reorientation ($\xi$). The model assumes that the MSMA is consisted of only two possible variants (Fig.~\ref{fig:7w}) $\xi_1$ and $\xi_2$, which as volume fraction must sum to 1:
\begin{equation}
\xi_1+\xi_2=1
\label{eqn:N1}
\end{equation}
and are subjected to the constraint:
\begin{equation}
0<\xi_i<1 
\label{eqn:N2}
\end{equation}
where \textit{i} = 1, 2. The other two internal variables shown in Fig.~\ref{fig:7w}, $\alpha_i$ and $\theta_i$ are independent of each other and bound by the following range:
\begin{eqnarray}
&0<\theta_i<\frac{\pi}{2}\\
\label{eqn:N3}
& 0<\alpha_i<1 
\label{eqn:N4}
\end{eqnarray}
The bounds on $\alpha_i$ are because it is also a volume fraction. The bounds on $\theta_i$ are based on physical constrain shown in Fig.~\ref{fig:7w}. When $\theta_i=0$ there is no rotation, and when $\theta_i=\frac{\pi}{2}$ the magnetization vector has fully rotate towards the hard axis in that variant.
With the offset angle $\beta$ added, the rotation of the internal magnetization vector of one of the variants will be changed, as the no field applied position of the internal magnetization vector changes. Fig. \ref{fig:7w}, shows a schematic of Cartesian component of the magnetization vector, which can be written as
\\
\begin{strip}
\begin{align}
\textbf{\textit{M}}=M_s\,\begin{Bmatrix}
\xi_2 \,\sin\left(\theta_2 \right)+\xi_1\, {\left(\alpha_1 -1\right)}\,\cos\left(\theta_1-\beta \right)\,+ \xi_1\, \alpha_1 \,\cos\left(\theta_1+\beta \right)\\
\xi_1 \,\alpha_1\,\sin\left(\theta_1+\beta \right)+\xi_1 \,(1-\alpha_1)\,\sin\left(\theta_1-\beta \right)+\xi_2\,\left(2\,\alpha_2 -1\right)\,\cos\left(\theta_2 \right)
\end{Bmatrix}
\label{eqn:N3_1}
\end{align}
\end{strip}
\\
where $M_s$ is the magnetization saturation value given by the manufacturer. Obeying the rules of thermodynamics and considering mechanical and magnetic energy sources, the Gibbs free energy takes the following form:
\begin{align}
\begin{split}
g\left(\boldsymbol{\sigma},\boldsymbol{H^{app}},\xi_i,\alpha_i,\theta_i\right)=\frac{1}{2\rho}\boldsymbol{\sigma}:\boldsymbol{S\sigma}-\frac{\mu_0}{\rho}\boldsymbol{M.H}\\+g^{an}+\frac{1}{\rho}f^\xi+\frac{1}{\rho}f^\alpha+\frac{1}{\rho}f^\theta
\label{eqn:N6}
\end{split}
\end{align}
where $\boldsymbol{\sigma}$ is stress,\textbf{\textit{ S}} is the compliance tensor, \textbf{\textit{M}} is the magnetization vector, \textbf{\textit{H}} is the magnetic field, $g^{an}$ is the anisotropy energy, and $f^{\xi}$, $f^{\alpha}$, $f^{\theta}$ are assumed hardening functions. These hardening functions capture how the evolution of the internal variables effect energy storage in the material. These terms aim to capture micro-structural interactions in a phenomenological sense. In this model, the values of $f^{\alpha}$ and $f^{\theta}$ are assumed to be zero. The $f^{\xi}$ term is used to help predict changes seen in the stress-strain and field-strain curves during reorientation.\\
The strain $\boldsymbol{\varepsilon}$ in MSMA can be expressed as
\begin{eqnarray}
\boldsymbol{\varepsilon}=\boldsymbol{\varepsilon^e}+\boldsymbol{\varepsilon^r}
\label{eqn:N7}
\end{eqnarray}
where $\boldsymbol{\varepsilon^e}$ is the elastic strain tensor and $\boldsymbol{\varepsilon^r}$ is the reorientation strain tensor.\\
The reorientation strain tensor $\boldsymbol{\varepsilon^r}$ can be written as
\begin{eqnarray}
\boldsymbol{\varepsilon^r}=\begin{bmatrix}
\xi_1 & 0\\
0 & \xi_2
\end{bmatrix}\varepsilon^{r,max}
\label{eqn:N8}
\end{eqnarray}
where $\varepsilon^{r,max}$ is the maximum achievable reorientation strain obtained from the experimental data.\\
The compliance tensor \textbf{\textit{S}} can be expressed as
\begin{eqnarray}
\boldsymbol{S}=\boldsymbol{S}^{\xi_1}\xi_1+\boldsymbol{S}^{\xi_2}\xi_2
\label{eqn:N9}
\end{eqnarray}
where $S^{\xi_1}$ and $S^{\xi_2}$ are the compliance tensors for variants 1 and 2, respectively.\\
Following~\cite{Keifer09, Alex12}, and neglecting the higher order term, the anisotropy energy $g^{an}$ \cite{Farrell2004} can be written as
\begin{eqnarray}
g^{an}=\xi_{1}\,K_1\,\sin^2\,\theta_1+\xi_2\,K_1\,\sin^2\,\theta_2
\label{eqn:N10}
\end{eqnarray}
where $K_1$ is the anisotropy coefficient which is the measurement of of the magnetization vectors’ resistance to rotation.\\
The internal magnetic field is assumed as follows:
\begin{equation}
\textbf{\textit{H}}=\textbf{\textit{H}}^{app}-\textbf{\textit{D}}\,\textbf{\textit{M}}
\label{eqn:N11}
\end{equation}
where \textbf{\textit{H}} is the internal magnetic field experienced by the MSMA specimen, $H^{app}$ is the external magnetic field applied on the specimen, \textbf{\textit{D}} is the demagnetization tensor \cite{Eberle_2019}.\\
Using Eq.~(\ref{eqn:N7}) $\&$ (\ref{eqn:N8}) in Eq.~(\ref{eqn:N6}), the Clausius-Duhem inequality can be written as
\begin{equation}
\begin{split}
\left(\sigma_{11}\,\varepsilon^{r,max}-\rho\frac{\partial g}{\partial \xi_1}\right)\dot{\xi_1}+\left(\sigma_{22}\,\varepsilon^{r,max}-\rho\frac{\partial g}{\partial \xi_2}\right)\dot{\xi_2}\\-\rho\frac{\partial g}{\partial \theta_1}\dot{\theta_1}-\rho\frac{\partial g}{\partial \theta_2}\dot{\theta_2}-\rho\frac{\partial g}{\partial \alpha_1}\dot{\alpha_1}-\rho\frac{\partial g}{\partial \alpha_2}\dot{\alpha_2}\,\geq\,0
\label{eqn:N12}
\end{split}
\end{equation}
Introducing $\pi$ as the driving force for change of each internal variable, Eq.~(\ref{eqn:N12}) can be written as
\begin{equation}
\pi^{\xi_1}\dot{\xi_1}+\pi^{\xi_2}\dot{\xi_2}+\pi^{\theta_1}\dot{\theta_1}+\pi^{\theta_2}\dot{\theta_2}+\pi^{\alpha_1}\dot{\alpha_1}+\pi^{\alpha_2}\dot{\alpha_2}\,\geq\,0
\label{eqn:N13}
\end{equation}
where the $\pi$ terms are defined as
\begin{gather}
\pi^{\xi_1}=\left(\sigma_{11}\,\varepsilon^{r,max}-\rho\frac{\partial g}{\partial \xi_1}\right)\,\,\\\pi^{\xi_2}=\left(\sigma_{22}\,\varepsilon^{r,max}-\rho\frac{\partial g}{\partial \xi_2}\right)\\\pi^{\theta_1}=-\rho\frac{\partial g}{\partial \theta_1},\,\,\pi^{\theta_2}=-\rho\frac{\partial g}{\partial \theta_2}\\\pi^{\alpha_1}=-\rho\frac{\partial g}{\partial \alpha_1},\,\,\pi^{\alpha_2}=-\rho\frac{\partial g}{\partial \alpha_2}
\end{gather}

In this model, $\alpha_i$ is piece-wise defined. When the applied field reaches a certain threshold, the value of alpha is considered 1, and less than that value of alpha is 0.5. This threshold value is called $H_{cri}^\alpha$, which is calculated from the M-H curve given by the manufacturer. The equation of alpha can be expressed as follows:
\begin{equation}
 \alpha_i= 
\begin{cases}
    0.5,& \text{if } H_i < H_{cri}^\alpha\\
    1,              & \text{otherwise}
\end{cases}
\label{eqn:N18}
\end{equation}
As $\alpha_i$ is piece-wise constant in the Eq.~(\ref{eqn:N18}), $\dot{\alpha_i}$ is zero almost everywhere. Thus, $\pi^{\alpha_i}\dot{\alpha_i}$ is also zero, which satisfies the thermodynamics. However, there is a possibility of the thermodynamics being violated at the single instant where $\alpha_i$ changes, but as it is only a single instant, this potential violation of thermodynamic requirements will be ignored.

Following~\cite{T_lance_2018}, the hardening function $f^\xi$ used in this model is
\begin{equation}
f^\xi=\frac{1}{6}C_1\,\xi_1^6+\frac{1}{2}C_2\,\xi_1^2
\label{eqn:N19}
\end{equation}
where $C_1$ \& $C_2$ are constants that need to be calibrated.

Finally, the model is completed by using following Kuhn-Tucker conditions to ensure positive dissipation during variant reorientation:
\begin{equation}
\begin{cases}
    \pi^{\xi_1}-\pi^{\xi_2}=Y, &  \dot{\xi_1} >0\\
    \pi^{\xi_1}-\pi^{\xi_2}=-Y, & \dot{\xi_1} <0
\end{cases}
\label{eqn:N22}
\end{equation}
where Y is a positive constant and calibrated from experimental data.

\subsection{Model Calibration}\label{cali}
\begin{figure}[htbp]
\centering
\subfloat[Experimental curve]{{\includegraphics[scale=0.25]{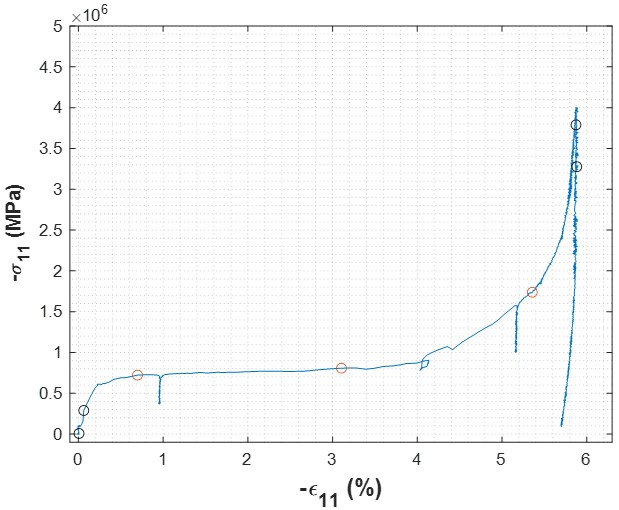}}}
\qquad
\subfloat[Predicted curve]{{\includegraphics[scale=0.25]{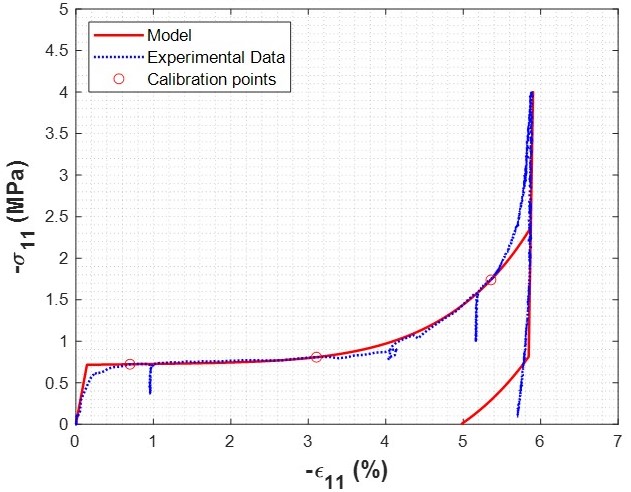}}}
\caption{Calibration test with uni-axial load and zero field, experimental stress-strain curve is on the left side (a), black dots are the points for calculating the compliance tensor and the red three dots on the reorientation region are the points for calculating the constants $C_1, C_2, Y$, predicted stress-strain curve (solid red line) is shown on the right side (b).}
\label{fig:10_9w}
\end{figure}

The model needs to be calibrated with the experimental data to find the hardening constants $C_1$, $C_2$, and the positive constant of the Kuhn-Tucker condition Y. Following~\cite{LaMaster2014}, the simplified loading condition is used for calibration. For calibration, the specimen is brought into a fully elongated state by applying a lateral field. Then the lateral field is removed, and the specimen is gradually compressed to 4.5 MPa no field applied. As there is no magnetic field, the magnetic domains $\alpha_i$ are 0.5, and there is no variant rotation, i.e., $\theta_i=0$. Using these values in Eq.~(\ref{eqn:N22}) and considering $\dot{\xi_1}>0$, as the specimen is transitioning from variant 2 to variant 1, the following equation is obtained:
\begin{equation}
Y+2\,C_1\,{\xi_1^5}_i+2\,C_2\, {\xi_1}_i=e^{r,max}{\sigma_{11}}_i+{\sigma_{11}^2}_i (S^{\xi_1}-S^{\xi_1})
\label{eqn:C1}
\end{equation}
where i=1, 2, 3; ${\sigma_{11}}_i$ and ${\xi_1}_i$ are the compressive stress and variant volume fraction of the $i-{th}$ point on the experimental stress-strain curve with zero field shown in Fig. \ref{fig:10_9w}(a).

\begin{figure}[htbp]
\centering
\includegraphics[scale=0.5]{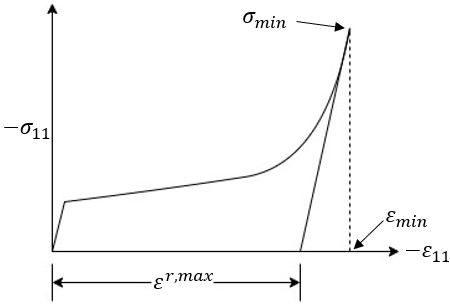}
\caption{Schematic of measuring maximum reorientation strain $\varepsilon^{r,max}$.}
\label{fig:28w}
\end{figure}

The variant volume fraction (${\xi_1}_i$) shown in Eqn.~(\ref{eqn:C1}) can be found from the experimental data, using the Eqns.~(\ref{eqn:N7}), (\ref{eqn:N8}) $\&$ (\ref{eqn:N9}), as
\begin{equation}
{\xi_1}_i=\frac{{\varepsilon_{11}}_i-S^{\xi_2}\,{\sigma_{11}}_i}{(S^{\xi_1}-S^{\xi_2}){\sigma_{11}}_i\,\varepsilon^{r,max}}
\label{eqn:C2}
\end{equation}
The compliance tensor $S^{\xi_i}$ is calculated by taking the inverse of the modulus of elasticity, which is measured directly from the experimental stress-strain curve. Two different moduli of elasticity are found; one is in variant 2 (fully elongated state), and the other is in variant 1 (fully compressed state) from the two distinct elastic regions in Fig.~\ref{fig:10_9w}(a). Specifically, the two points in each elastic region (see Fig. \ref{fig:10_9w}(a), black dots) are used to calculate the slope, which gives the modulus of elasticity. The maximum reorientation $\varepsilon^{r,max}$ is calculated from the experimental data by using the following equations:
\begin{equation}
\varepsilon^{r,max}=\varepsilon_{min}-\frac{\sigma_{min}}{E_1}
\label{eqn:C3}
\end{equation}
where $\varepsilon_{min}$ is the maximum negative strain of the data set, $\sigma_{min}$ is the maximum compressive stress of the experimental data, and $E_1$ is the modulus of elasticity of the variant 1. The origins of this equation are shown in Fig.~\ref{fig:28w}.
Finally, all the parameters of Eq.~(\ref{eqn:C1}) are known except $C_1$, $C_2$ $\&$ Y. These constants can be calculated by solving a system of three linear equations, which is obtained by taking three points on the calibration curve (see Fig.~\ref{fig:10_9w}(a), red dots). These calibrated constants are listed in Table~\ref{tab:T2}.

\begin{table}[htbp]
\caption{\label{tab:T1}Material Parameters.}
\centering
\begin{tabular}{ c   c } 
 \hline\hline
 $\rho\,K_1$ & \SI{0.89e5}{\joule\per\meter^3} \\ 
 $M_s$ & \SI{5.44e5}{\ampere\per\meter} \\ 
 ${H^\alpha}_{cri}$ & \SI{23000}{\ampere\per\meter} \\ 
 $\mu_0$ & 4$\pi$$\times10^7$ $Hm^{-1}$ \\
 $D_{11}$ & 0.0674 \\
 $D_{22}$ & 0.466 \\
 \hline\hline
\end{tabular}
\end{table}
\begin{table}[htbp]
\caption{\label{tab:T2}Calibrated Constants.}
\centering
\begin{tabular}{ c   c } 
 \hline\hline
 $C_1$ & 39408.25 \\ 
 $C_2$ & 9962.91 \\ 
 $Y$ & 44181.38 \\ 
 $S^{\xi_1}$ & \SI{1.957e-10}{\meter^2\per\newton} \\
 $S^{\xi_2}$ & \SI{2.247e-9}{\meter^2\per\newton} \\
 $e^{r,max}$ & -0.057867 \\
 \hline\hline
\end{tabular}
\end{table}
Using the material parameters and constants listed in Table~\ref{tab:T1} $\&$ \ref{tab:T2} and the derived model, the simulation is run, and the predicted calibration curve is shown in Fig. \ref{fig:10_9w}(b). As expected, the calibration curve is well matched by the simulation and the simulation exactly matches the experiment at the red points used for calibration.

\subsection{Introducing EMF to the Model}
From Faraday’s law of induction, the following equations can be used to calculate emf:
\begin{equation}
E=-NA\,\frac{d\,B}{d\,t}\\
\label{eqn:N24}
\end{equation}
\begin{equation}
B=\mu_0\left(H+M\right) 
\label{eqn:N25}
\end{equation}
where E is emf, N is the number of turns in the coil, A is cross-section area, $\mu_0$ is magnetic permeability constant, and t is time.

Using Eq.~(\ref{eqn:N11}), (\ref{eqn:N24}) \& (\ref{eqn:N25}), emf can be simplified to
\begin{equation}
E=-NA\,\mu_0\left(1-D_{11}\right)\frac{d\,M_1}{d\,t}
\label{eqn:N26}
\end{equation}
where $D_{11}$ is the 11 component of demagnetization tensor, and $M_1$ is the magnetization vector component in 1 direction shown in Fig.~\ref{fig:2w} obtained from the model. $\frac{d\,M_1}{d\,t}$ can be calculated numerically by assuming
\begin{equation}
\frac{d\,M_1}{d\,t}\simeq\frac{\Delta\,M_1}{\Delta\,t}
\label{eqn:N27}
\end{equation}
where $\Delta\,M_1$ is the difference in magnetization between two consecutive time steps, and $\Delta\,t$ is the time difference between those steps.

\section{Experiments}
The MSMA specimen under the experimental configuration shown in Fig.~\ref{fig:2w} was used to gather emf data from the coil. Several experiments were conducted by Nelson et al.~\cite{Nelson2014} and Guiel et al.~\cite{Guiel2018} to get voltage output from the MSMA specimen. Nelson et al.~\cite{Nelson2014} performed the stress-controlled test to be consistent with the LaMaster's model~\cite{LaMaster2014}. But instead of using the axial field, Nelson et al.~\cite{Nelson2014} tilted the lateral field at a certain angle. The tilted lateral field was not ideal because, in their setup, it was not possible to get a uniform magnetic field over the MSMA specimen. Moreover, more recent experimental result with a similar setup suggested that these experiments might have had significant variability of the stress and this input might not have been controlled as well as originally thought.

On the other hand, Guiel et al.~\cite{Guiel2018} performed the strain-controlled test, i.e., the Instron Machine was kept in position-controlled mode. In this setup, the maximum strain was measured experimentally before each cycle. But in this case, the specimen was experienced higher compressive stress ($\approx$ 8 MPa), which caused cracks in the specimen.

For several reasons these experiments could not be directly compared with model predictions. First, the Guiel et al. experiments were strain-controlled while the model is stress controlled. Next, the field was measured at the surface in these tests, which accordingly to Eberle et al.~\cite{Eberle_2019} should not be used as the input applied field. Finally, the Guiel et al. experiments, which were performed at 15 Hz frequency showed evidence that inertial effects were evident (i.e. the strain obtained from the tests at 15 Hz seemed different than the strain obtained from those tests done very slowly at 2 Hz).

Therefore, to be consistent with the model, additional stress-controlled experiments under the conditions shown in Fig.~\ref{fig:2w} were performed at a slower rate of 10 Hz in order to compare the results with the model. At this frequency level, the strain seems consistent with the strain obtained from the test of 2 Hz frequency. The power harvesting experimental data used to validate test this model was acquired through stress-controlled tests that were performed with improved PID control parameters, correcting the errors encountered in the data reported by Nelson et al.~\cite{Nelson2014}. During these tests, the axial compressive stress was varied between 0 to 5 MPa, while the specimen is exposed to constant axial and lateral fields.

\begin{figure}[H]
\centering
\includegraphics[scale=0.5]{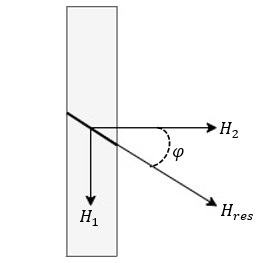}
\caption{Schematic of MSMA with magnetic field and angle. $H_2$ is the lateral field, $H_1$ is the axial magnetic field, $H_{res}$ is the resultant magnetic field, and $\phi$ is the angle between the lateral field and the resultant field.}
\label{fig:22w}
\end{figure}

As shown by Guiel et al.~\cite{Guiel2018}, the maximum emf can be obtained with a certain angle ($\phi$) of the resultant magnetic field with the horizontal axis (see Fig~\ref{fig:22w}). The angle $\phi$ is calculated by using the following equation:
\begin{equation}
\phi=\tan^{-1}\left(\frac{H_1}{H_2}\right)
\label{eqn:N3_1_1}
\end{equation}
The resultant field $H_{res}$ is calculated as:
\begin{equation}
H_{res}=\sqrt{H_1^2+H_2^2}
\label{eqn:N3_1_2}
\end{equation}
\FloatBarrier
The emf has been measured experimentally at several angles ranging from 0 to 7.2\si{\degree} while keeping the resultant field unchanged at 0.578 T. Using these angles and associated field values, simulations have been performed to calculate emf from the revised model.

\begin{figure}[H]
\centering
\includegraphics[scale=0.5]{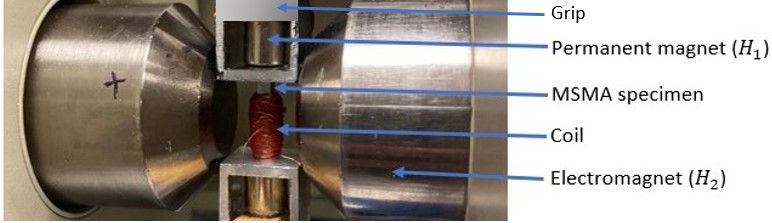}
\caption{Experimental setup for the power harvesting test. Axial field $H_1$ is applied on the specimen by using permanent magnet and lateral field $H_2$ is applied by using electromagnet. Different values of $\phi$ are obtained by varying $H_1$ $\&$ $H_2$. Note that $H_{res}$ is always kept same.}
\label{fig:24w}
\end{figure}

The experimental setup for power harvesting is shown in Fig.~\ref{fig:24w}. The axial field ($H_1$) was applied using the permanent magnets as shown in Fig.~\ref{fig:24w}, lateral field ($H_2$) was applied by using an electromagnet shown in Fig.~\ref{fig:24w}. The magnitude of these magnetic fields were measured by using a Hall probe and the angle was calculated by using Eq.~(\ref{eqn:N3_1_1}). The compressive stress was applied to the sample through the grip shown in Fig.~\ref{fig:24w} by using an 8874 Instron testing machine. The machine was kept in load controlled mode and the applied forces on the specimen were recorded via Instron Machine associated data acquisition software \say{Wavematrix 2}. The force applied on the specimen and the displacement of the top part of the sample were recorded automatically by the software controlling the machine. The strain could not be measured during the experiment with the available equipment because the coil surrounded the sample~(see Fig.~\ref{fig:24w}). The strain could be calculated, although not very accurately, from the position data collected. Before running the experiment, a white dot was placed on one of the sides of the specimen through which the lateral filed was applied. When the dot was kept in the top left side of the specimen, this configuration was assumed as $+\beta$, and when the dot was held at the top right side of the specimen, the configuration was assumed as $-\beta$. 
\FloatBarrier

\section{Generalized Regression Neural Network}

Generalized regression neural network (GRNN) which was proposed by D.F. Specht \cite{97934}, was adapted from radial basis neural networks \cite{BROOMHEAD1988} so that it can be used for regression, prediction, and classifications.
\begin{figure}[H]
\centering
\includegraphics[scale=0.5]{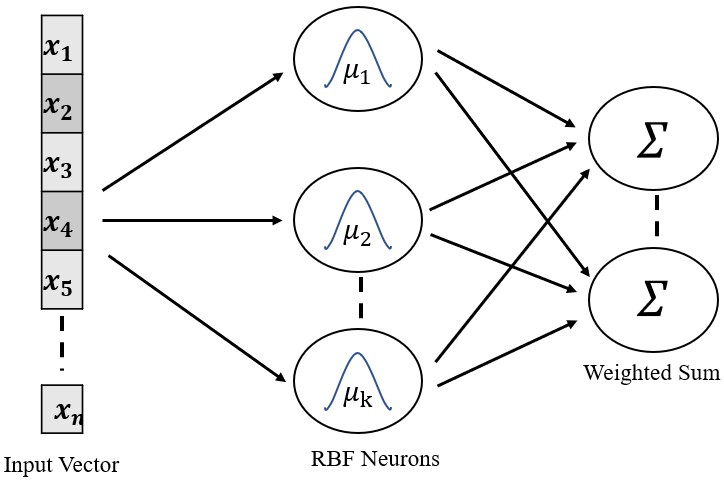}
\caption{Simplified architecture of Radial Basis Function Network (RBFN).Here \textbf{\textit{$x_n$}} is the input vector and $\mu$ is the prototype. The prototype vector may be defined as the center of the neuron since the prototype is the value of the center of the bell curve.}
\label{fig:30w}
\end{figure}
\FloatBarrier
GRNN represents an enhanced technique in the neural networks on the basis of non-parametric regression \cite{bowman1997applied} which bounds every training sample to represent a mean to a radial basis neuron. Mathematically it can be expressed as
\begin{equation}
Y(x)=\dfrac{\sum_{k=1}^{N}y_k K(x, x_k)}{\sum_{k=1}^{N}K(x, x_k)}
\label{eqn:D24}
\end{equation}
where $Y(x)$ is the predicted value of input $x$, $y_k$ is the activation weight for the pattern layer neuron at $k$, $K(x, x_k)$ is the Radial basis function kernel (RBFK) \cite{2549}. Here Gaussian kernel is used as RBFK which can be expressed as
\begin{equation}
K(x, x_k)=e^{-d_k/{2\sigma^2}}
\label{eqn:D25}
\end{equation}
where $d_k$ is the squared euclidean distance between the training samples ($x_k$) and the input ($x$) which is expressed as
\begin{equation}
d_k=(x-x_k)^T(x-x_k)
\label{eqn:D26}
\end{equation}
Eq. \ref{eqn:D24} can be modified as \cite{97934}:
\begin{equation}
E(F, H_1, H_2)=\dfrac{\sum_{k=1}^{N}A^k e^{-d_k/{2\sigma^2}}}{\sum_{k=1}^{N}B^k e^{-d_k/{2\sigma^2}}}
\label{eqn:D34}
\end{equation}
where E indicates the estimator of emf, $A^k$ $\&$ $B^k$ represent the coefficients for the cluster.
\begin{figure}[ht]
\centering
\includegraphics[scale=0.7]{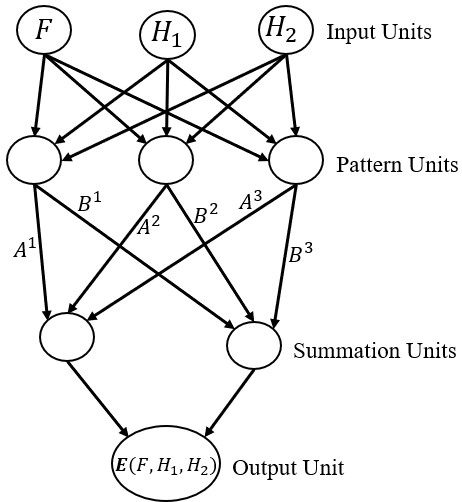}
\caption{Simplified architecture of Generalized Regression Neural Network (GRNN) for predicting emf.}
\label{fig:31w}
\end{figure}
\FloatBarrier
In this work, the GRNN is chosen for the prediction because of its advantages over other neural networks. In our study, the data set is small and GRNN is efficient working with less volume of data and it utilizes Gaussian functions which increase the prediction accuracy. Moreover, GRNN does not require any backpropagation to update the weights which makes it possible to avoid the limitation of backpropagation \cite{Lecun2015}. In Fig. \ref{fig:31w} a feed-forward neural network is shown which can be used to estimate the emf E from the measurement vector, force (F), and magnetic field (H).
\begin{figure}[H]
\centering
\includegraphics[scale=0.35]{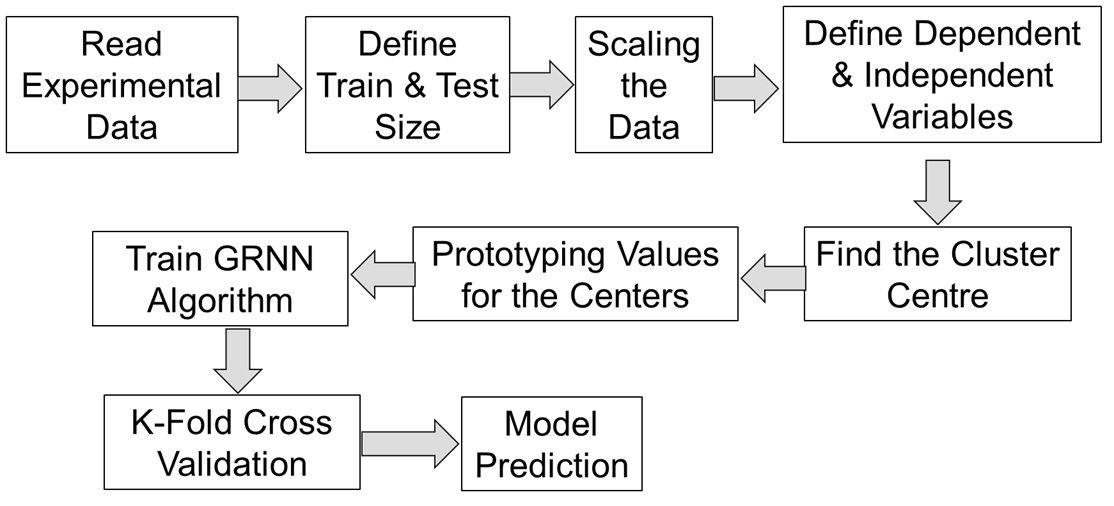}
\caption{Simplified Flow Chart of Machine Learning Based Modeling.}
\label{fig:24d}
\end{figure}
\FloatBarrier
In Fig. \ref{fig:24d}, a simplified workflow chart of the machine learning-based modelling is shown. We had a total of four variables, axial force ($F$), axial magnetic field ($H_1$), lateral magnetic field ($H_2$), and emf, among these three, are independent variables and one is the dependent variable (emf). 75$\%$ of the total data was used for training the model and the rest 25$\%$ was used to test the model using the builtin \say{test-train split} function of scikit-learn. All the values of the data set were scaled in-between 0 to 1 to increase the model efficiency. k-fold cross-validation is used in this model where the original sample is randomly separated into k equal-sized subgroups. Among k subgroups, a single group is picked as the validation data for testing the model, and the rest $(k-1)$ subgroups are used as training data.

\section{Results}
\begin{figure}[h]
\centering
\includegraphics[scale=0.8]{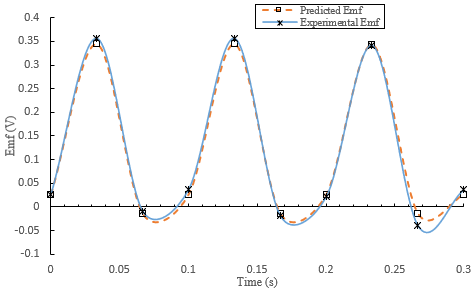}
\caption{Emf vs. time plots. The solid blue curve represents the experimental emf with positive $\beta$ while the dashed orange curve represents the model predicted emf with GRNN model. The lateral and axial magnetic fields were 0.575 \& 0.057 T, respectively.}
\label{fig:26d}
\end{figure}

The experimental results shown here were obtained using the loading frequency of 10 Hz.
\begin{figure}[H]
\centering
\includegraphics[scale=0.8]{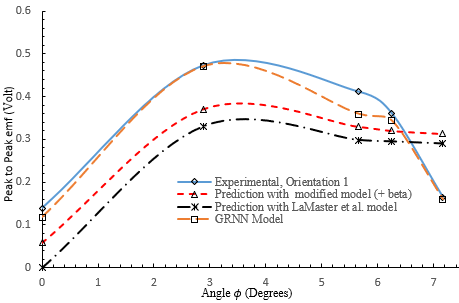}
\caption{Peak to peak emf vs. angle plots. The solid blue curve represents the experimental emf with positive $\beta$. The dashed red, black, \& orange curve represent the model predicted emf with modified model, LaMaster et. al. model \cite{LaMaster2014}, \& GRNN model, respectively.}
\label{fig:25d}
\end{figure}
\FloatBarrier
The predicted curve shown in Fig.~\ref{fig:25d} follows the same trend as the experiment, suggesting that this model reasonably predicts the angle for maximum emf. However, discrepancies between the model and experiment suggest that further improvement is needed in constitutive modeling. However, GRNN predicts the experimental curve very well. In Fig.~\ref{fig:26d}, experimental and machine learning model predicted emf is compared with a function of time. The GRNN model captures the experimental curve with a accuracy of 99$\%$.
\FloatBarrier

\section{Discussion}
The primary purpose of this study was to accurately predict the open circuit emf output of MSMA in power harvesting applications. This prediction can be used to design and optimize power harvesting devices. Towards that end, emf was predicted from the LaMaster et al.~\cite{LaMaster2014}, a modified model where an experimentally observed offset angle was added to the LaMaster et al. model, and generalized regression neural network (GRNN) model. All models were used to predict experimental emf data and the results are shown in Fig.~\ref{fig:25d}. The experimental emf shown in Fig.~\ref{fig:25d} \&\ref{fig:26d} were obtained by performing a load control test with a magnetic field applied at various angles. Note that the models all assume that the applied field was uniform, but that was hard to achieve experimentally.

\begin{figure}[H]
\centering
\includegraphics[scale=0.5]{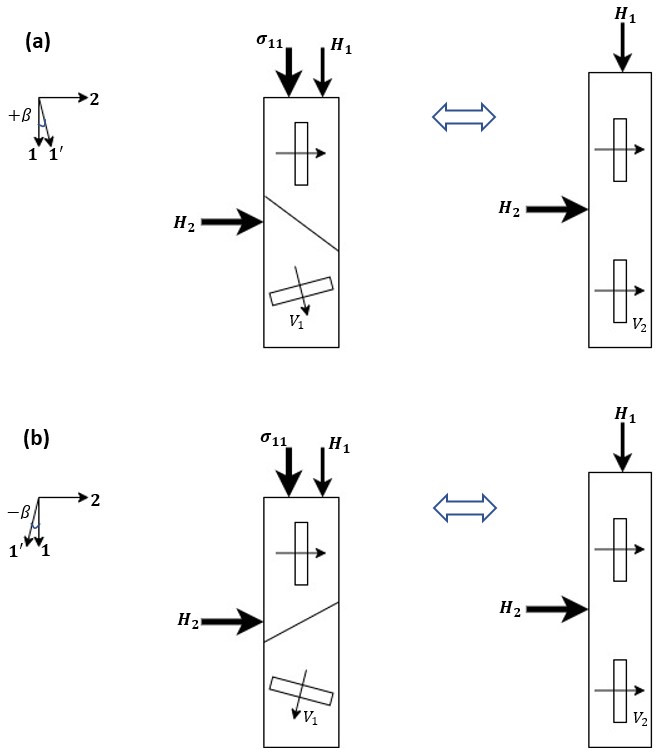}
\caption{Schematic of MSMA with two different orientations, i.e., positive $\beta$ (a) and negative $\beta$ (b). Constant lateral ($H_2$) and axial ($H_1$) magnetic field are applied on the specimen. Varying axial compressive stress ($\sigma_{11}$) is also applied. $V_1$ and $V_2$ represent stress preferred, and field preferred variants, respectively.}
\label{fig:8w}
\end{figure}

\begin{figure}[H]
\centering
\includegraphics[scale=0.3]{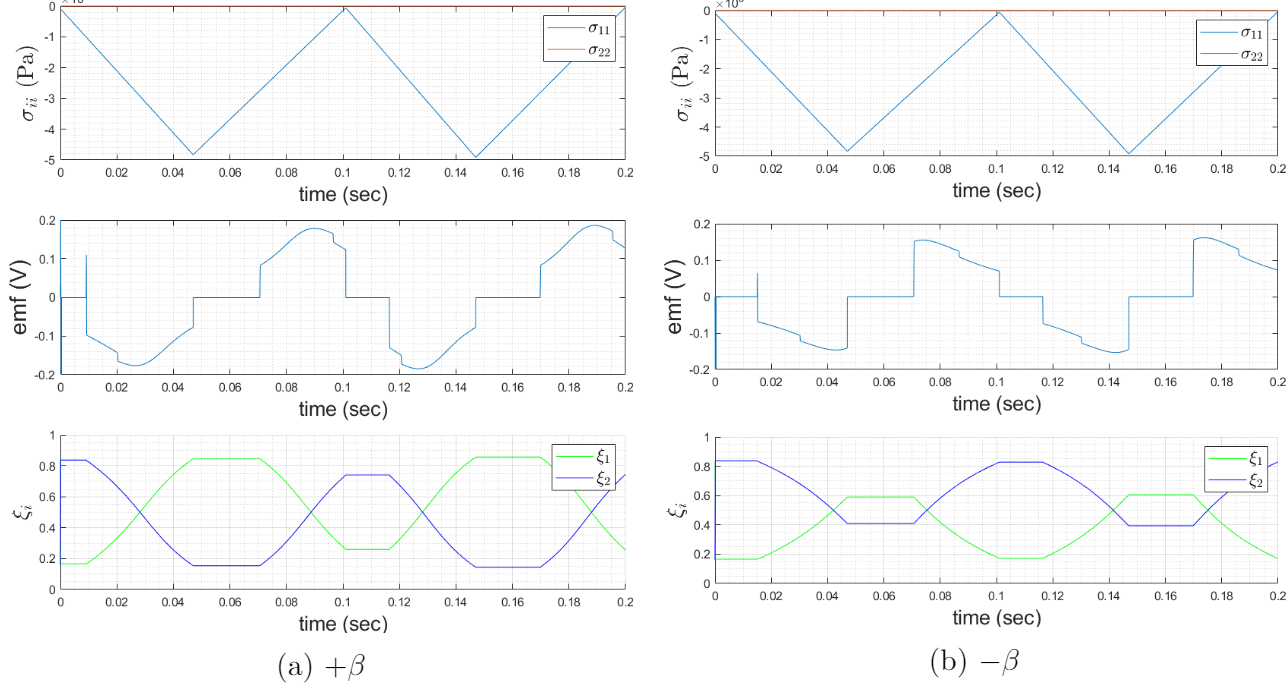}
\caption{Emf and variant volume fraction prediction with associate load condition from the model with offset angle added for both (a) positive and (b) negative $\beta$. Peak to peak voltage in (a) $\&$ (b) are 0.371 $\&$ 0.315 volts, respectively. Note that full reorientation is not obtained here due to low axial stress (0-4.8 MPa) and resultant field (0.578 T) level. The axial field $H_1$ is 0.029 T.}
\label{fig:11w}
\end{figure}

\begin{figure}[H]
\centering
\includegraphics[scale=0.3]{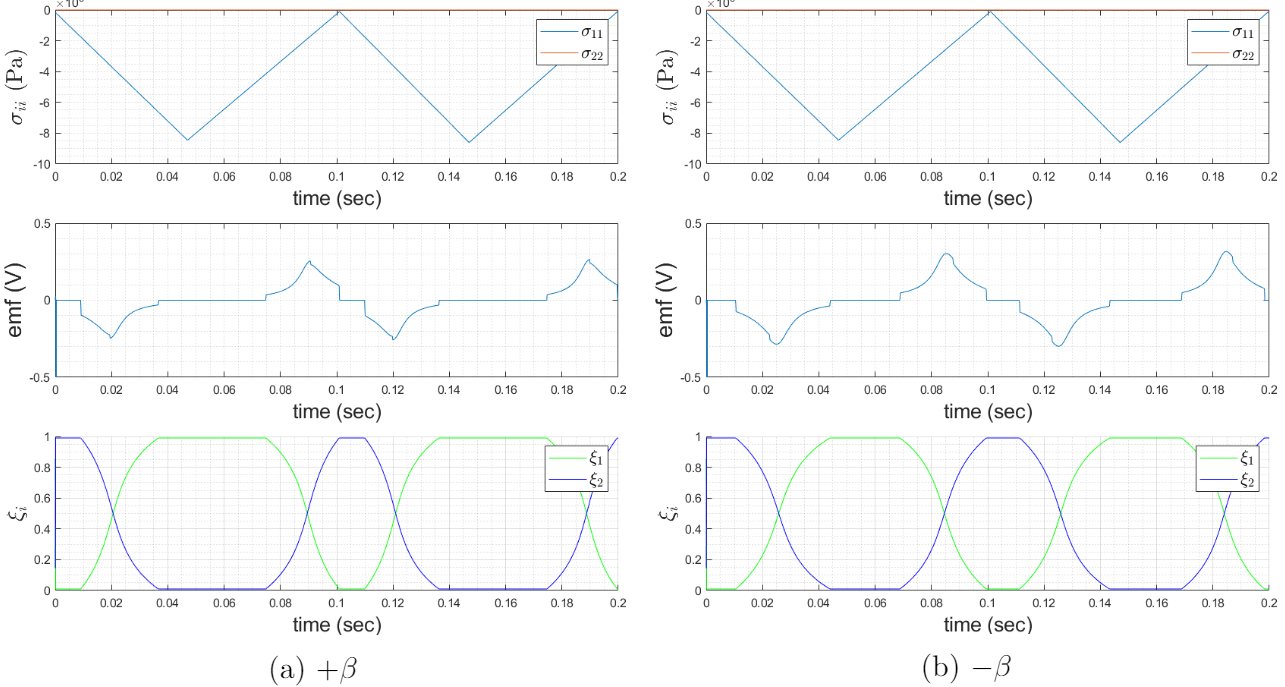}
\caption{Emf and variant volume fraction prediction with associate load condition from the model with offset angle added for both (a) positive and (b) negative $\beta$. Peak to peak voltage in (a) $\&$ (b) are 0.5131 $\&$ 0.6021 volts, respectively. Note that full reorientation is obtained here due to high axial stress (0-8.3 MPa) and resultant field (0.7 T) level. The axial field $H_1$ is 0.018 T.}
\label{fig:11w2}
\end{figure}

One of the major contributions of this study is to add an offset angle to the model which helps to explain some of the power-harvesting capabilities of MSMAs. Using Fig. \ref{fig:8w}, the reason for getting different peak to peak emf from different orientations~\cite{Nelson2014} of the MSMA can be hypothesized. There seem to be competing factors explaining how $\beta$ affects emf output. On the one hand, Fig.~\ref{fig:8w}(a) with $\beta$ positive seems to make reorientation easier. As the $\xi_i$ vs. time plot in Fig.~\ref{fig:11w} shows, there is more reorientation, and the slope is stiffer with $+\beta$ than with $-\beta$ because of the requirements of less energy to reorient. Thus, change in magnetization within a certain time period is more in Fig.~\ref{fig:11w}(a) than Fig.~\ref{fig:11w}(b). As the difference in magnetization is higher, so the emf is also higher (Eq.~(\ref{eqn:N26})). On the other hand, in Fig.~\ref{fig:8w}(b) there is more overall change in magnetization as the specimen changes from the stress preferred state to the field preferred state. Thus, when the full reorientation is obtained (see Fig.~\ref{fig:11w2}), $-\beta$ gives more emf than $+\beta$. In this case, the stress and field level is very high, more than 8 MPa and 0.7 T, respectively. As the variants are fully reoriented, there is no additional reorientation of the case of $+\beta$ and instead, the more change in magnetization in $-\beta$ case leads to more emf with $-\beta$.

\FloatBarrier

\FloatBarrier

\FloatBarrier
In this work, the experiments performed were stress-controlled, and the stress varies between 0 to 4.8 MPa. Therefore, full reorientation was not obtained. Thus, the ease of reorientation dominates, and as seen earlier, the $+\beta$ configuration (see Fig.~\ref{fig:8w}(a)) produces more emf than $-\beta$ configuration (see Fig.~\ref{fig:8w}(b)). The configuration which produces more emf experimentally than the other is assumed as $+\beta$ configuration in Fig.~\ref{fig:25d}. The idea that $-\beta$ produces more emf with very large fields and stresses was not verified experimentally, however, it should be as part of future work. The hypothesizes about the sign of $\beta$ and the maximum emf suggests that two different types of energy harvester are possible to design, one is for low stress level, and another one is for high stress level or strain-controlled. In either case, Fig.~\ref{fig:25d} suggests that the peak to peak emf is maximum at an angle between 3-7\si{\degree} and this power harvesting devices of the type in Fig.~\ref{fig:24w} should have a small axial field to maximize emf output. However, as the angle depends on the magnitude of the axial and lateral field, it is necessary to pick suitable magnets for maximizing the emf output.

In Fig.~\ref{fig:25d} the predicted emf obtained from the modified model (red dashed line) falls closer to the experimental emf than the LaMaster et al. model (black dashed line). This suggests that adding a new feature, i.e, offset angle to the model increases the model accuracy. However, there is still a significant difference between experimental and the predicted results which suggests that some key features are still missing from the constitutive model. Model calibration can be the possible reason behind the inconsistency of poor prediction. All the constitutive models were calibrated at zero axial fields while the experiments were performed with the presence of different axial field values. This may be the possible reason for the deviation of predicted trend extremely with increasing of the angle (Fig. \ref{fig:25d}).

In order to overcome all the constrain of constitutive modeling, a generalized regression neural network (GRNN) model is implemented in this study to make accurate predictions. GRNN model is extensively dependent on experimental data since the model learns from the data and finds the hidden co-relation. In our study, the GRNN model was able to predict the experimental curve with a high level of accuracy. Besides capturing peak to peak emf (Fig.~\ref{fig:25d}), GRNN model can predict emf as a function of time (Fig.~\ref{fig:26d}) exceptionally well.

\section{CONCLUSIONS}
Since the constitutive models in this work were not particularly accurate enough at predicting emf, their use in design is limited, and significant future work is needed to find the missing features. Some future work is needed to test some of the hypothesizes and understanding of the models in this work regarding the emf at the high-stress level. Further study is also recommended to find any other form of energy that may be included in Gibbs free energy to improve the prediction of variant reorientation. One such kind of energy is magnetic exchange energy which is the fundamental property of ferromagnetic materials that favors alignment of the magnetization vector spins along a common direction~\cite{Garcia}. Although constitutive modeling does not produce good predictions, the GRNN model was successful to predict the behavior and these machine learning approaches can be useful in designing MSMA-based sensors and power harvesters.

\addtolength{\textheight}{-10cm}   



\section*{ACKNOWLEDGMENT}
This work has been supported by the National Science Foundation under Grant No. 1101108 and Grant No.~1561866. Any opinions, findings, and conclusions or recommendations expressed in this article are those of the author(s) and do not necessarily reflect the views of the National Science Foundation. The author would like to thank Glen D’Silva, for his constructive feedback and helping him doing the experiments. The author would also like to thank Professor Sybil Derrible for helping him to implement the machine learning algorithm.

\bibliographystyle{ieeetr}
\bibliography{main}

\begin{thebibliography}{10}

\bibitem{Silva}
G.~J.~D. Silva, C.~Ciocanel, and H.~P. Feigenbaum, ``Visualization of magnetic
  domains and magnetization vectors in magnetic shape memory alloys under
  magneto-mechanical loading.,'' {\em Shape memory and superelasticity},
  vol.~6, March 2020.

\bibitem{Eberle_2019}
J.~L. Eberle, H.~P. Feigenbaum, and C.~Ciocanel, ``Demagnetizing field in
  single crystal ferromagnetic shape memory alloys,'' {\em Smart Materials and
  Structures}, vol.~28, p.~025022, jan 2019.

\bibitem{Guiel2018}
R.~Guiel, H.~Feigenbaum, and C.~Ciocanel, ``{The effect of magnetic field
  orientation on the open-circuit voltage of Ni-Mn-Ga based power
  harvesters},'' {\em Smart Materials and Structures}, vol.~27, jul 2018.

\bibitem{Minorowicz_2016}
B.~Minorowicz, G.~Leonetti, F.~Stefanski, G.~Binetti, and D.~Naso, ``Design,
  modelling and control of a micro-positioning actuator based on magnetic shape
  memory alloys,'' {\em Smart Materials and Structures}, vol.~25, p.~075005,
  may 2016.

\bibitem{Ullakko2012}
K.~Ullakko, L.~Wendell, A.~Smith, P.~M{\"{u}}llner, and G.~Hampikian, ``{A
  magnetic shape memory micropump: Contact-free, and compatible with PCR and
  human DNA profiling},'' {\em Smart Materials and Structures}, vol.~21,
  no.~11, 2012.

\bibitem{Nelson2014}
I.~Nelson, J.~Dikes, H.~Feigenbaum, and C.~Ciocanel, ``{Numerical predictions
  versus experimental findings on the power-harvesting output of a NiMnGa
  alloy},'' {\em Behavior and Mechanics of Multifunctional Materials and
  Composites 2014}, vol.~9058, no.~March 2014, p.~905815, 2014.

\bibitem{Kiefer2005}
B.~Kiefer and D.~C. Lagoudas, ``{Magnetic field-induced martensitic variant
  reorientation in magnetic shape memory alloys},'' {\em Philosophical
  Magazine}, vol.~85, no.~33-35, pp.~4289--4329, 2005.

\bibitem{Wang2012}
J.~Wang and P.~Steinmann, ``{A variational approach towards the modeling of
  magnetic field-induced strains in magnetic shape memory alloys},'' {\em
  Journal of the Mechanics and Physics of Solids}, vol.~60, no.~6,
  pp.~1179--1200, 2012.

\bibitem{LaMaster2014}
D.~H. LaMaster, H.~P. Feigenbaum, I.~D. Nelson, and C.~Ciocanel, ``{A full
  two-dimensional thermodynamic-based model for magnetic shape memory
  alloys},'' {\em Journal of Applied Mechanics, Transactions ASME}, vol.~81,
  no.~6, pp.~1--12, 2014.

\bibitem{DesignedResearch;M2019}
M.~Mozaffar, R.~Bostanabad, W.~Chen, K.~Ehmann, J.~Cao, and M.~A. Bessa,
  ``{Deep learning predicts path-dependent plasticity},'' {\em Proceedings of
  the National Academy of Sciences of the United States of America}, vol.~116,
  no.~52, pp.~26414--26420, 2019.

\bibitem{Cyron2021}
G.~A. Holzapfel, K.~Linka, S.~Sherifova, and C.~J. Cyron, ``{Predictive
  constitutive modelling of arteries by deep learning},'' {\em Journal of the
  Royal Society Interface}, vol.~18, no.~182, 2021.

\bibitem{Linka2021}
K.~Linka, M.~Hillg{\"{a}}rtner, K.~P. Abdolazizi, R.~C. Aydin, M.~Itskov, and
  C.~J. Cyron, ``{Constitutive artificial neural networks: A fast and general
  approach to predictive data-driven constitutive modeling by deep learning},''
  {\em Journal of Computational Physics}, vol.~429, p.~110010, 2021.

\bibitem{Zhang2022}
Z.~Zhang, Q.~Liu, and D.~Wu, ``{Predicting stress–strain curves using
  transfer learning: Knowledge transfer across polymer composites},'' {\em
  Materials and Design}, vol.~218, p.~110700, 2022.

\bibitem{97934}
D.~Specht, ``A general regression neural network,'' {\em IEEE Transactions on
  Neural Networks}, vol.~2, no.~6, pp.~568--576, 1991.

\bibitem{Tellinen2002}
J.~Tellinen, I.~Suorsa, I.~Aaltio, and K.~Ullakko, ``{Basic Properties of
  Magnetic Shape Memory Actuators},'' {\em 8th international conference
  ACTUATOR 2002}, pp.~10--12, June 2002.

\bibitem{niskanen2013}
A.~J. Niskanen and I.~Laitinen, ``Design and simulation of a magnetic shape
  memory (msm) alloy energy harvester,'' in {\em State-of-the-Art Research and
  Application of SMAs Technologies (4th CIMTEC)}, vol.~78 of {\em Advances in
  Science and Technology}, pp.~58--62, Trans Tech Publications Ltd, January
  2013.

\bibitem{T_lance_2018}
J.~L. Eberle, ``Predicting the magneto-mechanical behavior of single crystal
  magnetic shape memory alloys using homogenized models.,'' Master's thesis,
  Northern Arizona University, 2018.
\newblock ProQuest13423996.

\bibitem{Keifer09}
B.~Kiefer and D.~Lagoudas, ``Modeling the coupled strain and magnetization
  response of magnetic shape memory alloys under magnetomechanical loading,''
  {\em Journal of Intelligent Material Systems and Structures}, vol.~20,
  pp.~143--170, 01 2009.

\bibitem{Alex12}
A.~Waldauer, H.~Feigenbaum, C.~Ciocanel, and N.~Bruno, ``Improved thermodynamic
  model for magnetic shape memory alloys,'' {\em Smart Materials and Structures
  - SMART MATER STRUCT}, vol.~21, 09 2012.

\bibitem{Farrell2004}
S.~P. Farrell, R.~A. Dunlap, L.~M. Cheng, R.~Ham-Su, M.~A. Gharghouri, and
  C.~V. Hyatt, ``{Magnetic properties of single crystals of Ni-Mn-Ga magnetic
  shape memory alloys},'' {\em Smart Structures and Materials 2004: Active
  Materials: Behavior and Mechanics}, vol.~5387, no.~July 2004, p.~186, 2004.

\bibitem{BROOMHEAD1988}
D.~BROOMHEAD, ``{Multivariable functional interpolation and adaptive
  networks},'' {\em Complex Systems}, vol.~2, pp.~321--355, 1988.

\bibitem{bowman1997applied}
A.~Bowman and A.~Azzalini, {\em Applied Smoothing Techniques for Data Analysis:
  The Kernel Approach with S-Plus Illustrations}.
\newblock Oxford Statistical Science Series, OUP Oxford, 1997.

\bibitem{2549}
J.~Vert, K.~Tsuda, and B.~Sch{\"o}lkopf, {\em A Primer on Kernel Methods},
  pp.~35--70.
\newblock Cambridge, MA, USA: MIT Press, 2004.

\bibitem{Lecun2015}
Y.~Lecun, Y.~Bengio, and G.~Hinton, ``{Deep learning},'' {\em Nature},
  vol.~521, no.~7553, pp.~436--444, 2015.

\bibitem{Garcia}
C.~J. García-Cervera, ``Numerical micromagnetics: A review,'' 2007.

\end{thebibliography}
\end{document}